\documentclass[preprintnumbers,superscriptaddress,endnote,nofootinbib,aps,prd,floatfix,12 pt]{revtex4}
\pdfoutput=1
\usepackage[utf8x]{inputenc}     
\usepackage{enumerate}
\usepackage{epsfig} 
\usepackage{float} 
\usepackage[colorlinks = true,
linkcolor =blue,
urlcolor  = blue,
citecolor = blue,
anchorcolor = blue]{hyperref}
\usepackage[caption = false]{subfig}
\usepackage{wasysym} 
\usepackage{mathrsfs} 
\usepackage{amsfonts} 
\usepackage{amsbsy} 
\usepackage{amscd} 
\usepackage{pstricks} 
\usepackage{multirow} 
\usepackage{tikz}
\usepackage{color}
\usepackage{slashed}
\usepackage{array}
\usepackage{tablefootnote}
\usepackage{url} 
\usetikzlibrary{arrows,positioning,shapes.geometric} 
\usepackage[compat=1.1.0]{tikz-feynman}          
\usepackage[font=small,labelfont=bf,
            justification=raggedright,
            format=plain]{caption}
\usepackage{slashed}
\usepackage{multirow}
\usepackage{tabularx}
\usepackage{soul}  
\usepackage{listings} 
\usepackage{color}

\begin{document}
\title{Influence of QCD parton shower in deep learning\\ invisible Higgs through vector boson fusion}
\flushbottom

\author{Partha Konar}\email{konar@prl.res.in}
\affiliation{Physical Research Laboratory, Ahmedabad - 380009, Gujarat, India\\[0.1cm]}

\author{Vishal S. Ngairangbam}\email{vishalng@prl.res.in}
\affiliation{Physical Research Laboratory, Ahmedabad - 380009, Gujarat, India\\[0.1cm]}
\affiliation{Indian Institute of Technology, Gandhinagar - 382424, Gujarat, India}

\begin{abstract} 
	Vector boson fusion established itself as a highly reliable channel to probe the Higgs boson and an avenue to uncover new physics at the Large Hadron Collider. This channel provides the most stringent bound on Higgs' invisible decay branching ratio, where the current upper limits are significantly higher than the one expected in the Standard Model. It is remarkable that merely low-level calorimeter data from this characteristically simple process can improve this limit substantially by employing sophisticated deep-learning techniques. The construction of such neural networks seems to comprehend the event kinematics and radiation pattern exceptionally well. However, the full potential of this outstanding capability also warrants a precise theoretical projection of QCD parton showering and corresponding radiation pattern. This work demonstrates the relation using different recoil schemes in the parton shower with leading order and higher-order computation.
\end{abstract}

\maketitle

\newpage
\tableofcontents
\newpage
\section{Introduction}

The discovery of the last missing piece~\cite{CMS:2012qbp,ATLAS:2012yve} of the Standard Model (SM) of particle physics opened up a plethora of independent searches at the CERN Large Hadron Collider (LHC). Apart from looking for new physics signatures, it is paramount to verify whether the scalar is ``the Higgs" boson or its properties has some nonstandard nature~\cite{CMS-PAS-HIG-20-003,CMS:2021rsq,CMS-PAS-HIG-20-007,ATLAS:2020ior,ATLAS:2021jbf,ATLAS:2020kdi} different from those predicted in the SM. With no direct evidence of new physics so far, there's still scope that it can show up in the Higgs boson's physical properties. Hence the precise determination of Higgs's intrinsic property and coupling with other fundamental particles can provide subtle hints towards new physics. 

Although with a relatively lower cross-section, Vector Boson Fusion (VBF) channel~\cite{Cahn:1983ip,first_vbf, PhysRevD.47.101,PhysRevD.48.5162, Rainwater:1997dg} has long been advocated for a cleaner alternative in hardon colliders among different Higgs production mechanisms. Eventually, this channel has also shown promise in several new physics searches~\cite{Eboli:2000ze,Konar:2003pn,Konar:2006qx,Choudhury:2003hq,Datta:2001cy,Datta:2001hv,Datta:2000ja}.
The VBF channel provides the best channel for constraining the Higgs' invisible decays~\cite{Eboli:2000ze}, providing a strong bound in several dark matter~\cite{Han:2016gyy,Heisig:2019vcj,Arcadi:2019lka,Arcadi:2021mag,Bhattacherjee:2021rml,Argyropoulos:2021sav} scenarios. It is particularly important since the current upper limit of 11\%~\cite{ATLAS:2020kdi} is still significantly larger than the one expected in the SM ($\lesssim 0.1\%$ with the decay chain $h\to ZZ^*\to \nu\bar{\nu}\;\nu'\bar{\nu}'$). In general, the VBF channel relies on the production of colour-neutral heavy states through the collision of electro-weak gauge bosons radiated from initial partons and are always associated with two hard forward jets tagged as its characteristic signature. In the absence of coloured particle exchange between two parton lines, very little QCD jet activity~\cite{first_vbf,PhysRevD.47.101} is seen in the detector's central part, in the region between two forward tagged jets. The decay products of the heavy state (in the present case, the Higgs) is expected to be in this region, retaining the colour quietness. These characteristic features are well studied and greatly exploited to identify VBF processes over various QCD backgrounds, exhibiting a different jet formation pattern. In this respect, different high-level variables are constructed as proxies for these features. Among them, rapidity gap, central jet veto, the invariant mass of two forward jets and Zeppenfeld variable ($z^*_{j_3}$) are some of the crucial ones. The precise theoretical prediction~\cite{Han:1992hr,Figy:2003nv,Dreyer:2016oyx,Liu:2019tuy} of additional jet formation patterns is as important as the experimental measurements. This precise prediction is even more significant in a data-driven paradigm where different characteristics of both signal and background distributions are minutely encapsulated to determine the decision boundary.

The application of machine-learning algorithms has shown immense promise at the Large Hadron Collider (LHC)~\cite{Radovic2018,Guest:2018yhq,Bourilkov:2019yoi,Kim:2019wns,Amacker:2020bmn,deOliveira:2015xxd}. They have significantly improved the performance compared to variables motivated by our knowledge of physics. However, this comes at the cost of a reduced understanding, which is even more profound when we use low-level data to train Deep Neural Networks (DNNs). Moreover, training with simulated data can result in the networks learning features specific to the imperfect simulation, not present in real experimental data. The use of low-level inclusive event information for searches brings a few unique challenges, particularly due to the higher systematic uncertainties associated with the simulations. Given the superior ability of deep-learning algorithms to extract features, it is necessary to circumspect the effect of using simulations which better resemble the actual physics.   

It is well known that VBF processes receive modest corrections from higher-order QCD corrections~\cite{Han:1992hr,Figy:2003nv,Dreyer:2016oyx,Liu:2019tuy}. Using the full event information in an input-image to a Convolutional Neural Network (CNN) for VBF searches has shown a promising avenue~\cite{Ngairangbam:2020ksz}, where the CNN exploits the lack of central-jet activity in VBF processes. We note that it is essential to scrutinize the differences in leading order (LO) and next-to-leading (NLO) order simulations: the presence of a third jet needs the proper introduction of real and virtual corrections to the tree level process. Another issue of central importance in the simulation of VBF events is the inability of a global-recoil scheme in initial state radiations (ISR) of the parton showering algorithm to describe the central-jet activity correctly~\cite{Cabouat:2017rzi}. This study systematically investigates these issues for the VBF signal search with CNNs, taking the invisible decay as a proxy. Although the preceding arguments apply to the VBF production of weak-bosons, we presently ignore its effects as they are much lesser in proportion ($\sim5\%$ of the total background for the cuts used here). We also neglect the contribution of the Gluon-fusion events in the signal. The Global recoil scheme correctly produces the leading logarithmic behaviour, already incorporated in our previous analysis. A precise determination of its various effects demands a very high level of sophistication, requiring much higher perturbative and logarithmic accuracy. Although the cuts used in the analysis have a sizable amount of gluon-fusion contribution, the large amount of data from high luminosity LHC runs will provide ways to do precision analysis with more stringent cuts, with negligible contribution from gluon-fusion events.  These do not impede our final goal, as our intention is not to project experimental sensitivities but to usher pragmatism and careful examination while using inclusive event information as inputs to DNNs.

	Although DNNs generally perform better than ML algorithms utilising high-level variables, their usability in phenomelogical analyses is determined by our ability to simulate subtle aspects of the data accurately. To this end, we show the possibility of CNNs learning inaccurate representations of inclusive events due to a global recoil used in the simulation of VBF events. We find that
	\begin{itemize}
		\item The training performance is greatly reduced when we use signal simulated with a global-recoil recoil scheme on parton level events generated with leading order or next-to-leading accuracy and improves for a dipole recoil, with events generated at next-to-leading order accuracy showered with a dipole-recoil having the highest training accuracy.
		\item For each set of signal simulations, the highest validation accuracy is achieved for the network that used the same type during the training process with the same trend as the training accuracies. However, the validation performance of the NLO events showered with dipole recoil (which is the most accurate description of the actual events amongst the four signals used) is affected mildly by the kind of data used during the training. 		
	\end{itemize}
	Our findings show that CNNs can learn the underlying differences between VBF type events and the dominant QCD backgrounds, even when trained on sub-optimal simulated data.

The rest of this paper is organized as follows. In Section \ref{sec:ML_VBF}, we outline the CNN based improvement over the existing study and point out the significance of QCD radiation for better accuracy. Following Section \ref{sec:qcd_exp} discusses the parton shower scheme and NLO effects in the simulation of VBF Higgs signal. In Section \ref{sec:result}, we examine the impact of the different signal simulations on the trained network output and its performance. We conclude in Section~\ref{sec:conclusion}.

\section{Deep learning invisible Higgs produced via VBF}
\label{sec:ML_VBF}
In this section, we summarise the VBF search of the invisible Higgs decays proposed in ref.~\cite{Ngairangbam:2020ksz} using deep learning. Since the present study aims to scrutinize the dependence of deep-learning algorithms with low-level inputs on the simulation, we focus on the analysis using Convolutional Neural Networks (CNNs) with tower images as the input. 
\subsection{Data simulation and selection criteria}
\label{sec:data_sim} 
The background class consists of non-VBF and VBF type production of $Z$ and $W$ bosons with at least two hard jets, with the $Z$ boson decaying to neutrinos, and the $W$ to a charged lepton and a neutrino (contributes when the lepton fails the identification criteria). The parton-level events are generated using {\tt MadGraph5\_aMC@NLO}~\cite{Alwall:2014hca} (v2.6.5) at  13 TeV LHC. {\tt Pythia8}~\cite{Sjostrand:2014zea} (v8.243) is used to shower these events in the default global-recoil scheme for the Initial State Radiation(ISR). We match parton level cross-sections of processes where the additional jets arise from QCD vertices via the MLM procedure~\cite{Mangano:2002,Mangano:2006rw} (up to four jets for $Z$ and two for $W$). The showered events were passed through {\tt Delphes3}~\cite{deFavereau:2013fsa} (v3.4.1) for simulating a parametrized response of the CMS detector.  Jets of radius $R=0.5$ and $p_T>20$ are clustered with the anti-$k_t$~\cite{Cacciari:2008gp} algorithm implemented in the {\tt Fastjet}~\cite{Cacciari:2011ma} package. 

To have a standard benchmark, we followed the experimental search ref.~\cite{CMS:2018yfx} and replicated their shape-analysis\footnote{This study did not use any ML techniques for the final analysis, recent analyses~\cite{ATLAS:2020fry,ATLAS:2020tlo} on VBF searches have used ML techniques with high-level variables. However, searches using low-level detector information has not been performed for VBF searches to the best of our knowledge.} on our simulated data.
Except for the cut on the missing transverse momentum ($\slashed{E}_T>200$ GeV instead of $\slashed{E}_T>250$ GeV), we implemented the same baseline selection criteria in the shape analysis, which is summarised in Appendix~\ref{app:base_sel}. Importantly, the analysis used weaker cuts on the rapidity gap between the two tagging jets $|\Delta \eta_{jj}|>1$, and the dijet invariant mass $m_{jj}>200$ GeV, than is normally done in VBF searches, and had no central-jet veto. We therefore expect a significant contribution from gluon-fusion production of the Higgs with two additional jets, along with an increase in selected backgrounds. Thus, the signal class consisted of VBF and gluon-fusion contributions in ref.~\cite{Ngairangbam:2020ksz}, decaying to two invisible dark matter particles. The weaker cuts without any central-jet veto retain a larger fraction of the VBF signal, which can be segregated using powerful deep-learning techniques. The background and signal classes were formed by combining the different processes according to their expected proportions after the baseline selection criteria.  We used 100k training and 25k validation events for each category. The background class consists of 51.221\% and 44.896\% of non-VBF $Z$ and $W^\pm$ backgrounds, respectively, and 2.295\% and 1.587\% of VBF-origin $Z$ and $W^\pm$ backgrounds, respectively. 

\subsection{Input representation and preprocessing} 
\label{sec:prepro}
\begin{figure}[t]
	\centering	
	\includegraphics[scale=0.25]{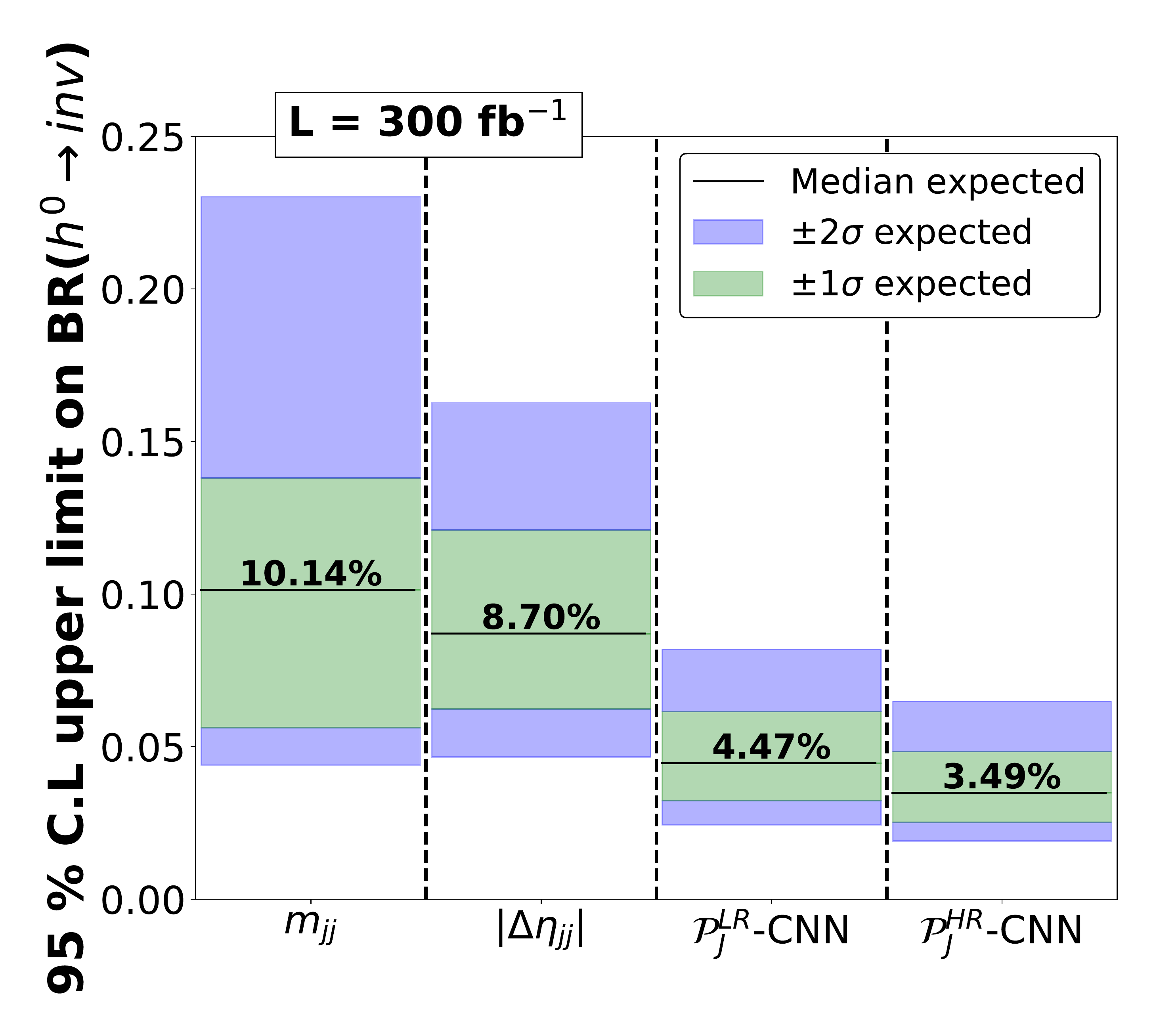}			
	\caption{Expected upper limits on the invisible branching ratio of the Higgs boson, for an integrated luminosity $L=300 \text{ fb}^{-1}$ from the analysis in ref.~\cite{Ngairangbam:2020ksz}.
		}		
	\label{fig:br_lim}
\end{figure} 
Although the analysis consisted of various high-level and low-level input representations with different preprocessing steps, we concentrate on the low-level representation with better regularised preprocessing steps. We first apply the following operations to the four-momenta of each particle in each event to regularize their spatial orientations:
\begin{enumerate}
	\item Rotate along z-axis such that the leading jet's center resides on the z-axis ($\phi_{j_1}=0$). 
	\item Reflect along the xy-plane, such that the leading jet's $\eta$ is always positive. 
\end{enumerate}
As inputs to CNNs, we form the tower-image of the full calorimeter in the $(\eta,\phi)$ plane, with the transverse energy $E_T$ as the pixel values, with two resolutions: $0.08\times0.08$ and $0.17\times0.17$. 
%
It might look like the leftovers due to imperfect divisibility by the pixel resolution from the full range of $\phi$ can contribute to a boundary effect. However, it is not detrimental to a CNN's performance since the pooling operations would effectively wash away its impact. Moreover, the images formed after the preprocessing sets $\phi_{j_1}=0$ further reduce the importance of the boundary by concentrating all useful information at the center. Periodicity of the $\phi$-axis is enforced by padding these images with a fixed number of rows from the opposite side.
Therefore, we get preprocessed high and low resolution images represented by $\mathcal{P}^{HR}_J$ and $\mathcal{P}^{LR}_J$, respectively. 

\subsection{Network architecture and training }
\label{sec:net_train}
The networks for the input images: $\mathcal{P}^{HR}_J$-CNN and $\mathcal{P}^{LR}_J$-CNN, has three blocks of sequential convolution-pooling operations. Each block consists of two convolutions with sixty-four $4\times4$ filters and an average pooling layer with a $2\times2$ pool size. The flattened output of the final pooling layer is fed to a dense network with three hidden layers containing 300 nodes each and an output layer with two nodes with a softmax activation for binary classification. All convolutional layers and the hidden dense layers have {\tt ReLu} activation. The network was trained for twenty epochs with {\tt Nadam}~\cite{Dozat2016IncorporatingNM} optimizer with a learning rate of 0.001, with cross-entropy loss and a batch size of 300. All training were implemented using {\tt Keras} (v2.2.4)~\cite{chollet2015keras} with a {\tt TensorFlow} (v1.14.1)~\cite{tensorflow2015-whitepaper} backend.

The exclusion limits on the invisible branching ratio of the Higgs boson obtained from the network output was compared to the shape analysis of $m_{jj}$ and $|\Delta \eta_{jj}|$. The expected upper limits for an integrated luminosity $L=300$ fb$^{-1}$ of the four scenarios are shown in Figure~\ref{fig:br_lim}. These limits were obtained using the CL$_s$ method \cite{Junk:1999kv,Read_2002} in the asymptotic approximation \cite{Cowan:2010js}, with the {\tt RooStats}~\cite{Moneta:2010pm} package. The statistical model was built using {\tt HistFactory}~\cite{Cranmer:2012sba}. The uncertainty bands are obtained by incorporating the per-bin statistical uncertainty, and the normalisation uncertainties of the total cross-section, Monte Carlo simulation, and the integrated luminosity. 
	
The CNN-based approach has better performance, putting stricter limits on the branching ratio. Moreover, it was found that training and validation on events with pile-up did not impact the performance and upper-bounds on the branching ratio considerable, increasing it mildly within the one-sigma errors obtained from the one without pile-up. Our study~\cite{Ngairangbam:2020ksz} did not apply any pileup mitigation technique on the pileup contaminated events and was trained and validated with tower-images contaminated with pileup. With extensive research and progress into process independent pileup contamination procedures, we expect this pileup contribution to reduce further. The lower transverse energy in the secondary pileup collisions significantly affects the forward regions. In contrast, for VBF processes, the parton-shower recoil schemes primarily involve the soft radiation in the central rapidity gap between the two (hard) forward jets. Moreover, as it will become clear from the results, CNNs primarily look at the amount of hard radiation between the two jets. Therefore, pileup effects would not conceal the impact of a third hard jet in VBF events, thereby demanding an accurate description of such hard jets, which is achieved up to leading-order in a next-to-leading order simulation of the complete process. Therefore, it is imperative to scrutinize these factors when dealing with deep-learning algorithms like CNNs, which enhance the physics reach by setting more stringent bounds on the invisible branching ratio of the Higgs boson by utilizing the inclusive event information efficiently.


\section{Impact of NLO corrections and recoil schemes} 
\label{sec:qcd_exp}
Although VBF processes have relatively lower higher-order corrections, utilising the hadronic activity between the two tagging jets would use information not captured by a leading order simulation. This inadequacy is due to the inherent assumption in parton shower generators, primarily focusing on the soft and collinear regions. A next-to-leading-order hard partonic simulation merged with a parton shower algorithm would accurately describe the kinematics of the third leading jet (if present) over the full range of transverse momentum. Additionally, for event topologies with no colour flow between the two incoming partons from the colliding protons, a parton shower algorithm with a global-recoil scheme for the initial-state radiation (ISR) is known to have a further inefficient simulation of the wide-angle soft radiation patterns. The cause for this inaccuracy is due to the incorrect assumption of an {\tt II} dipole in the global-recoil scheme~\cite{Cabouat:2017rzi}, while VBF processes have a double DIS scattering topology with an {\tt IF/FI} dipole structure. Existing phenomenological studies~\cite{Ballestrero:2018anz,Jager:2020hkz} are consistent with this known limitation of the global-recoil scheme, and recent experimental results~\cite{CMS:2021qzz,CMS:2021yqw,CMS:2020tkr} have employed the dipole recoil scheme~\cite{GUSTAFSON1986453,GUSTAFSON1988746,Schumann:2007mg,Platzer:2009jq} for the relevant VBF topologies.  The effects of both higher-order virtual corrections and the recoil scheme are even more important when using powerful deep-learning algorithms with low-level inputs.

\subsection{Signal generation}
 \label{sec:sig_gen}
 Since VBF Higgs processes are our primary interest, we do not include the gluon-fusion processes in the present analysis. We, therefore, study the four different possible combinations of the perturbative accuracy and the parton shower's recoil scheme for the VBF channel. These are described as follows:
  \begin{enumerate}
 	\item \textbf{Global-LO}: Parton level events simulated at leading-order perturbative accuracy showered with a global recoil scheme for the ISR parton shower. This recoil scheme is the default implementation in {\tt Pythia8} and was used in ref.~\cite{Ngairangbam:2020ksz} for the VBF processes. 
 	\item \textbf{Dipole-LO}: Parton level events simulated at leading-order perturbative accuracy showered with a dipole recoil scheme. 
 	\item \textbf{Global-NLO}: Parton level events simulated at next-to-leading order accuracy merged with parton shower employing the global recoil scheme for ISR.
 	\item \textbf{Dipole-NLO}: Parton level events simulated at next-to-leading order accuracy merged with parton shower using the dipole recoil scheme.
 \end{enumerate} 
 \begin{figure}[t]
 	\centering	
 	\includegraphics[scale=0.27]{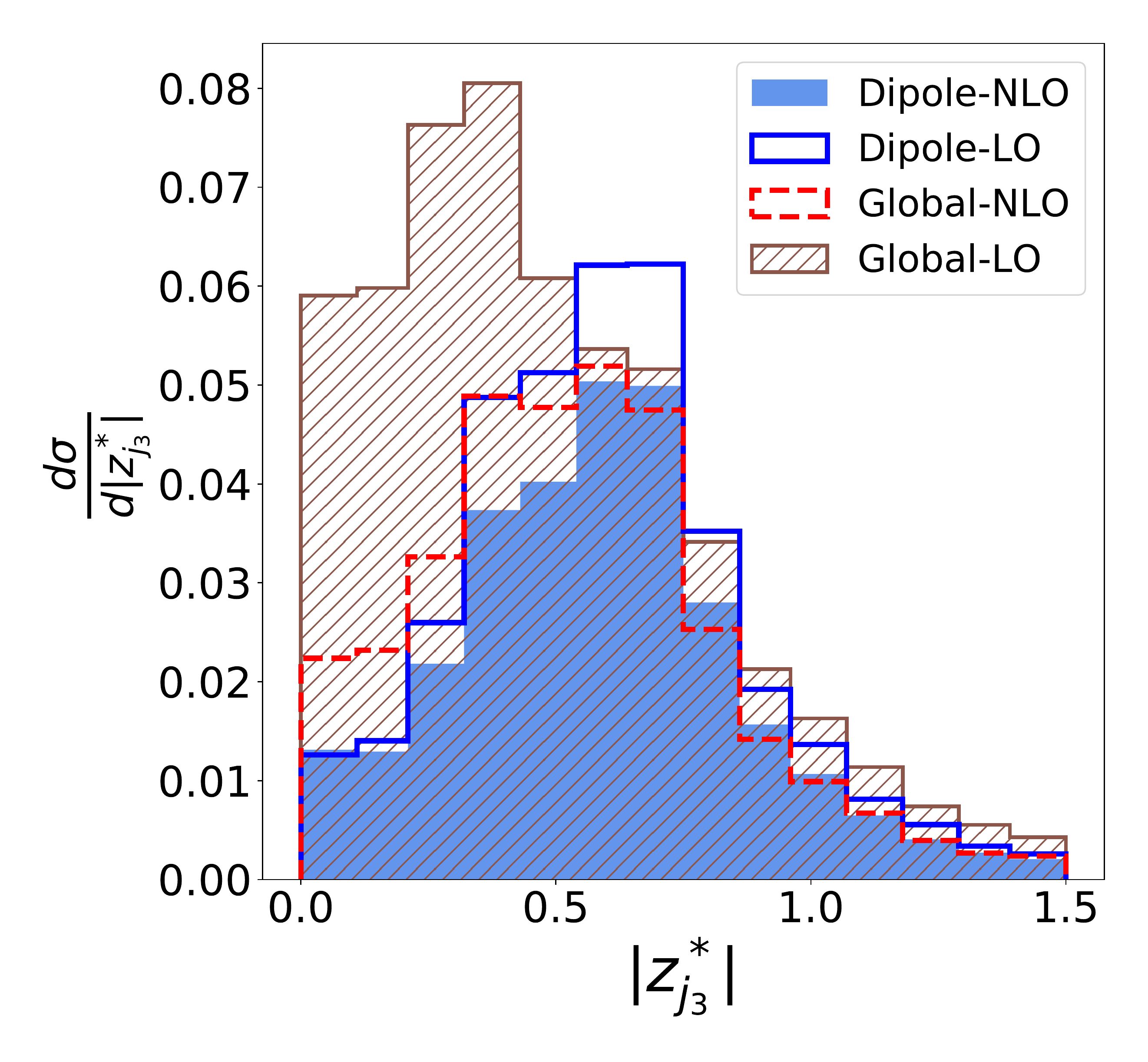}			
 	\caption{Distribution of the absolute Zeppenfeld variable $|z^*_{j_3}|$ for the four signal simulations. To capture the relative occurrence of the third jet, we set each event weight so that the total sum with or without an additional jet in each signal simulation sum to unity. }		
 	\label{fig:zep}
 \end{figure} 					
 
\begin{figure*}[t]
	\centering	
	\includegraphics[scale=0.2]{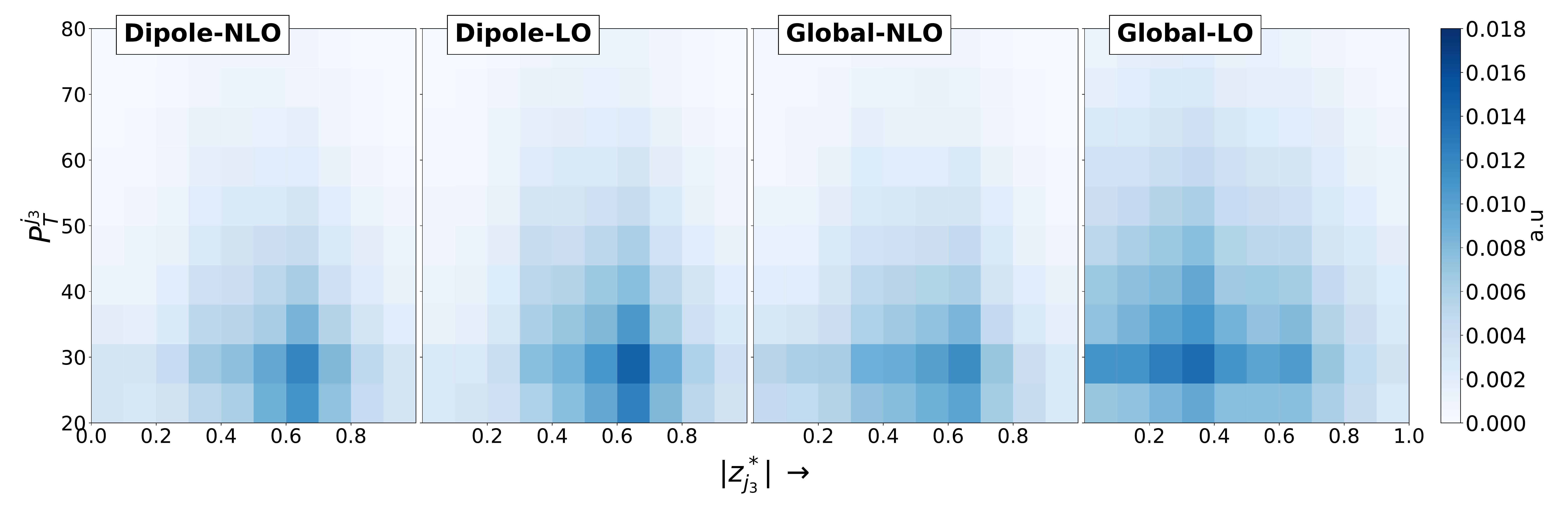}		
	\caption{Two dimensional histogram of events with the transverse momentum($P^{j_3}_T$) of the third jet and $|z^*_{j_3}|$ for four different cases of signal simulations, such as, dipole NLO, dipole LO , global NLO and global LO of the VBF Higgs signal.
	}		
	\label{fig:hist_2d}
\end{figure*}

 We use the same set of parton-level events for the LO and NLO simulations to shower with the two recoil schemes. The parton level events at LO were generated with  {\tt MadGraph5\_aMC@NLO},
  while the NLO events were generated with the {\tt POWHEG-BOX}~\cite{Nason:2004rx,Frixione:2007vw,Nason:2009ai,Alioli:2010xd}. The renormalization and factorization scales for both orders are set for each event as,
 \begin{equation}
 \label{eq:ren_fact}
 \mu_0=\frac{m_h}{2}\sqrt{\left(\frac{m_h}{2}\right)^2+p_{T,h}^2}\quad,
 \end{equation} where $m_h$ =125 GeV is the mass of the Higgs and $p_{T,h}$ is the transverse momentum of the Higgs boson in the event. For the parton level generation, we use the {\tt PDF4LHC15\_nlo\_100\_pdfas}~\cite{Butterworth:2015oua} parton distribution function (PDF) set implemented with {\tt LHAPDF6}~\cite{Buckley:2014ana} (v6.1.6) package. This PDF set is a combination~\cite{Forte:2010dt} of  {\tt CT14}~\cite{Dulat:2015mca}, {\tt MMHT14}~\cite{Harland-Lang:2014zoa}, and {\tt NNPDF3.0}~\cite{NNPDF:2014otw} PDF sets using the Hessian reduction method proposed in ref.~\cite{Carrazza:2015aoa}. We use {\tt MadSpin}~\cite{Artoisenet:2012st} to decay the Higgs boson at parton level to two scalar dark matter particles for the NLO events, while we simulate the full decay chain for the LO events. All parton showers are performed in {\tt Pythia8.235}. For the NLO events, we perform the powheg-merging with recommended values from ref.~\cite{pow_pythia}. The switch to a dipole-recoil scheme is done by setting {\tt "SpaceShower:dipoleRecoil=on"} for the parton shower.  We note that the events generated at NLO and showered with the dipole-recoil scheme should be the most physically accurate simulation of the VBF higgs process. These four sets of showered events are then passed through the same detector simulation and selection criteria\footnote{The details of the baseline selection are given in Appendix~\ref{app:base_sel}.} described in Section~\ref{sec:data_sim}.  We divide the dataset of each of these simulations into 100k training and 25k validation samples for the neural network analysis.

\subsection{Characteristics of the third jet}
\label{sec:third_char}

To compare the different signal simulations, we plot distributions of the Zeppenfeld variable $z^*_{j_3}$ in Figure \ref{fig:zep} for events passing the selection criteria and having a third jet with $p_T>20$ GeV. It is defined as,
\begin{equation}
\label{eq:zep}
z^*_{j_3}=\frac{\eta_{j_3}-(\eta_{j_1}+\eta_{j_2})/2}{|\Delta \eta_{j_1j_2}|}\quad,
\end{equation}  where $\eta_{j_i}$ is the pseudorapidity of the $i^{th}$ hardest jet, and $\Delta \eta_{j_1j_2}$ is the rapidity gap between the two tagging jets. This variable looks at the position of the third jet relative to the tagging jets and is important when considering the additional information available beyond the two jet system.  We set the normalization such that the cumulative sum of the bins correspond to the fraction of events that satisfy the requirement on the third jet. The Dipole-NLO signal has the least proportion of events passing the additional criteria at $30\%$, while the global NLO has $35\%$. The fraction for LO events with dipole and global recoil schemes are $37\%$ and 
$55\%$ respectively. From these values and the shape of the distribution in Figure \ref{fig:zep}, we can infer that out of the four, Global-LO should be most similar to the QCD dominated background, and Dipole-NLO should be the least identical. Consequently, we expect these to be reflected on the performance of any statistical model utilising radiative information beyond the two jets. Although the proportion of events with a third jet is very close for Global-NLO and Dipole-LO, note that the former has more jets in the central regions from the shape of $|z^*_{j_3}|$ distribution. Hence, we would expect better discrimination for Dipole-LO. 

Even though $z^*_{j_3}$ is a good variable, a model like a CNN that uses the inclusive event information will use the third jet's position as well as its transverse momentum implicitly to find the decision boundary. To this end, in Figure \ref{fig:hist_2d} we plot the 2-D histogram plot of the transverse momentum $P^{j_3}_T$ of the third jet and   $|z^*_{j_3}|$. Due to the artificial enhancement from the {\tt II} like global showering scheme in the central regions, we can see that the third jet is relatively harder than their dipole counterparts for both orders. Moreover, since the third jet results from the parton shower for LO, there is a drastic difference between Global-LO and Dipole-LO relative to the same comparison at NLO.  From this, we can infer that events that do not have a third reconstructed jet would still follow the same pattern and expect the same effect on the performance of the CNN.

%

\section{Results}
\label{sec:result}

\begin{figure*}
	\centering	
	\subfloat[]{\label{fig:out_pt}\includegraphics[scale=0.2]{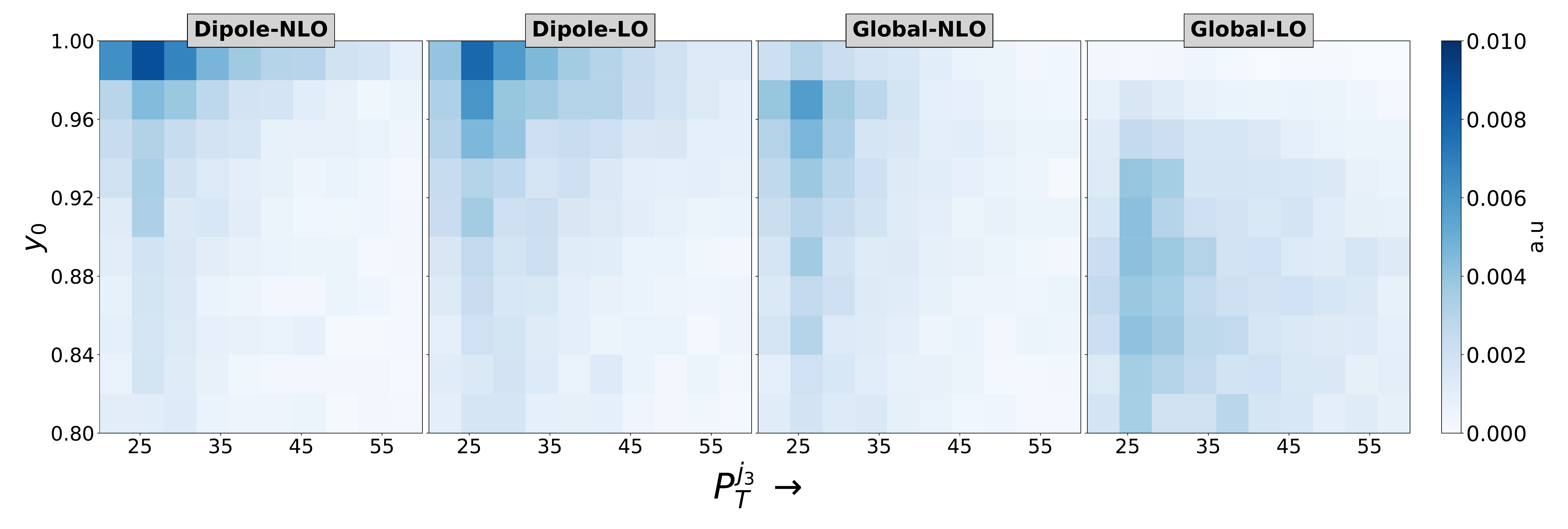}}\\
	\subfloat[]{\label{fig:out_eta}\includegraphics[scale=0.2]{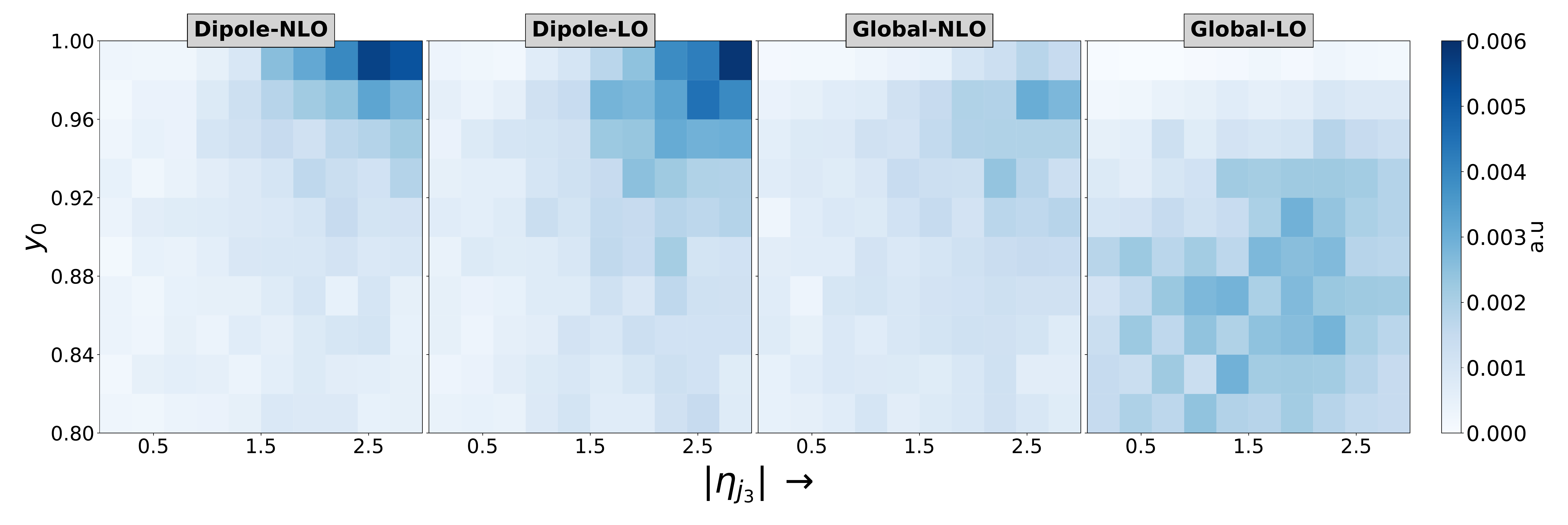}}\\	
	\subfloat[]{\label{fig:out_htc}\includegraphics[scale=0.2]{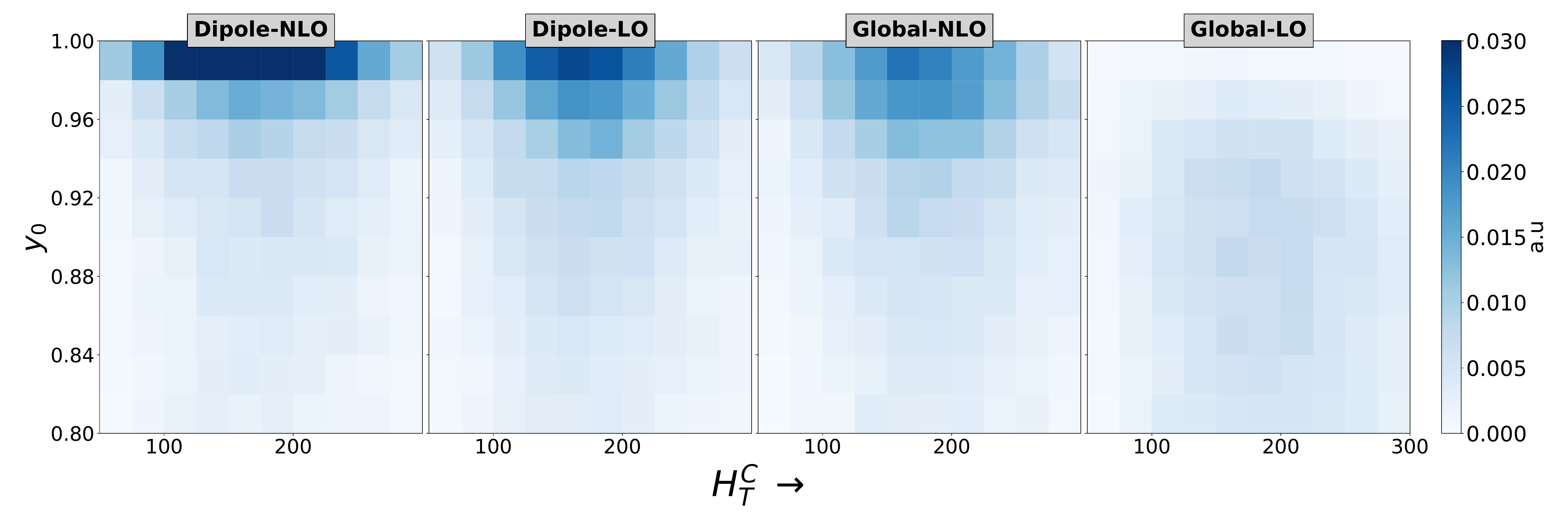}}		
	\caption{Two dimensional histogram of events of the network output $y_0$ for each signal simulation with the (a) $P^{j_3}_T$ and (b) $|\eta_{j_3}|$  of the third jet (when present), and (c) the $H^C_T$ between the two tagging jets.}		
	\label{fig:out_2d}
\end{figure*} 

\begin{figure*}[t!]
	\centering	
	
	\includegraphics[scale=0.18]{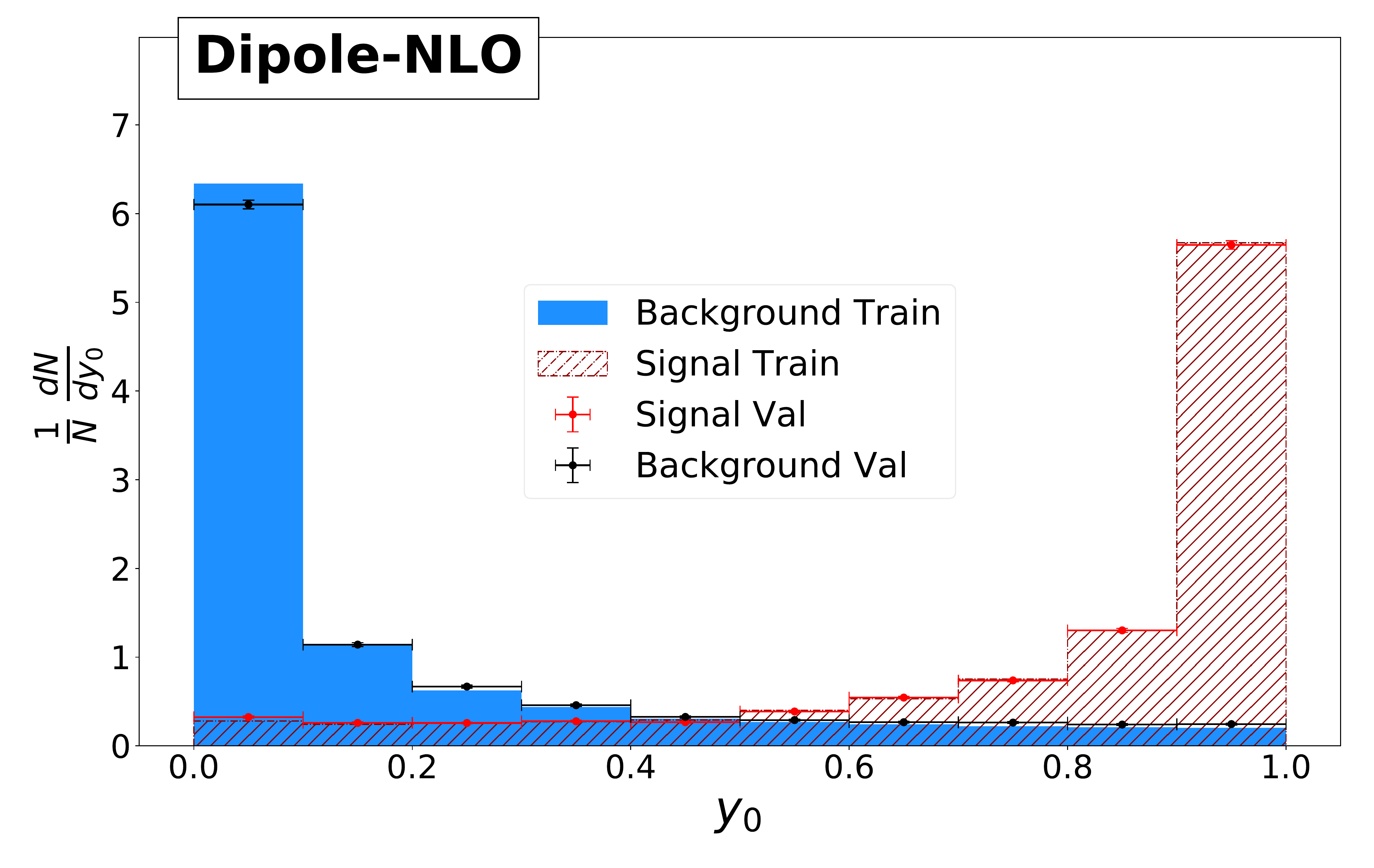}
	\includegraphics[scale=0.18]{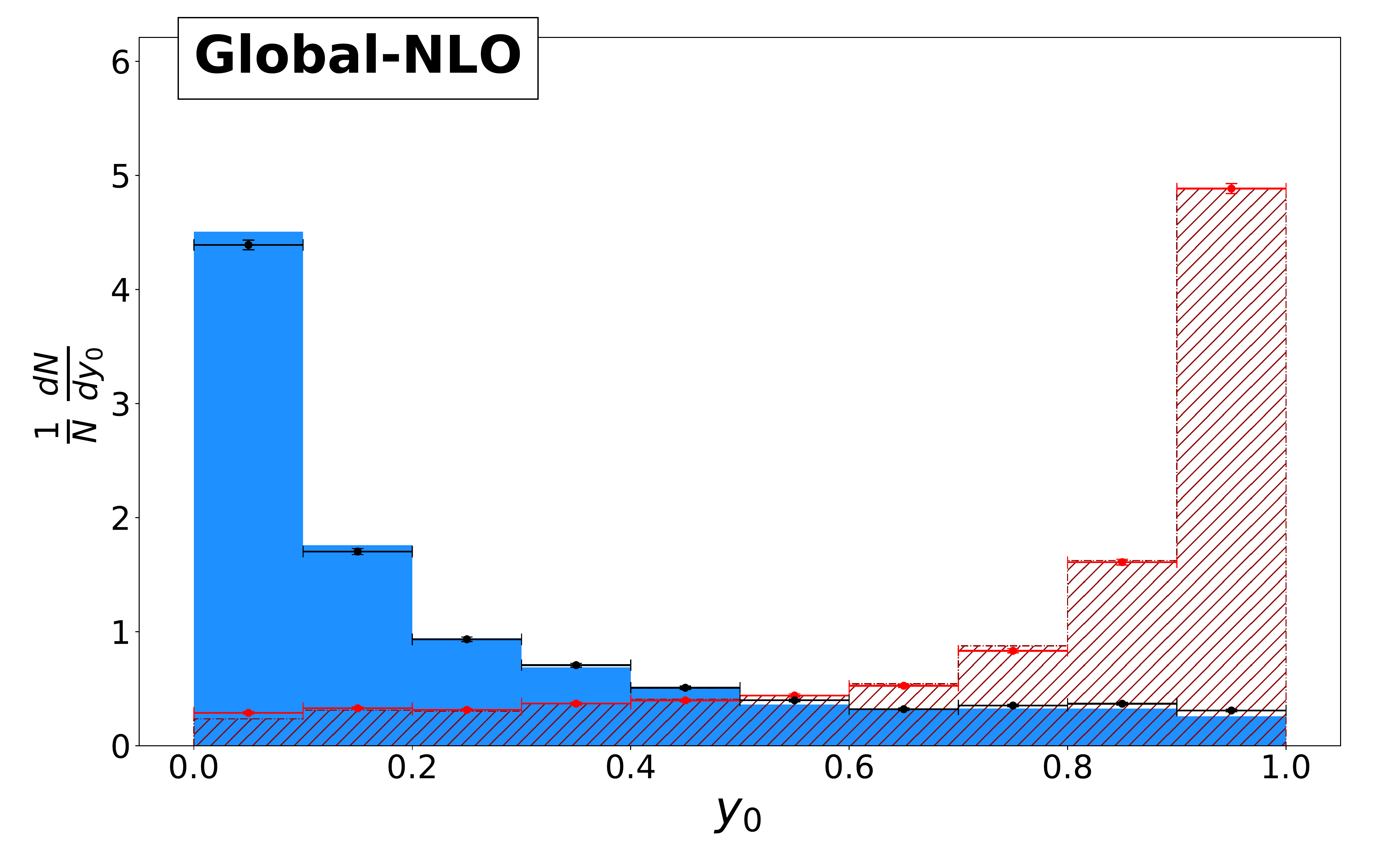}\\
	\includegraphics[scale=0.18]{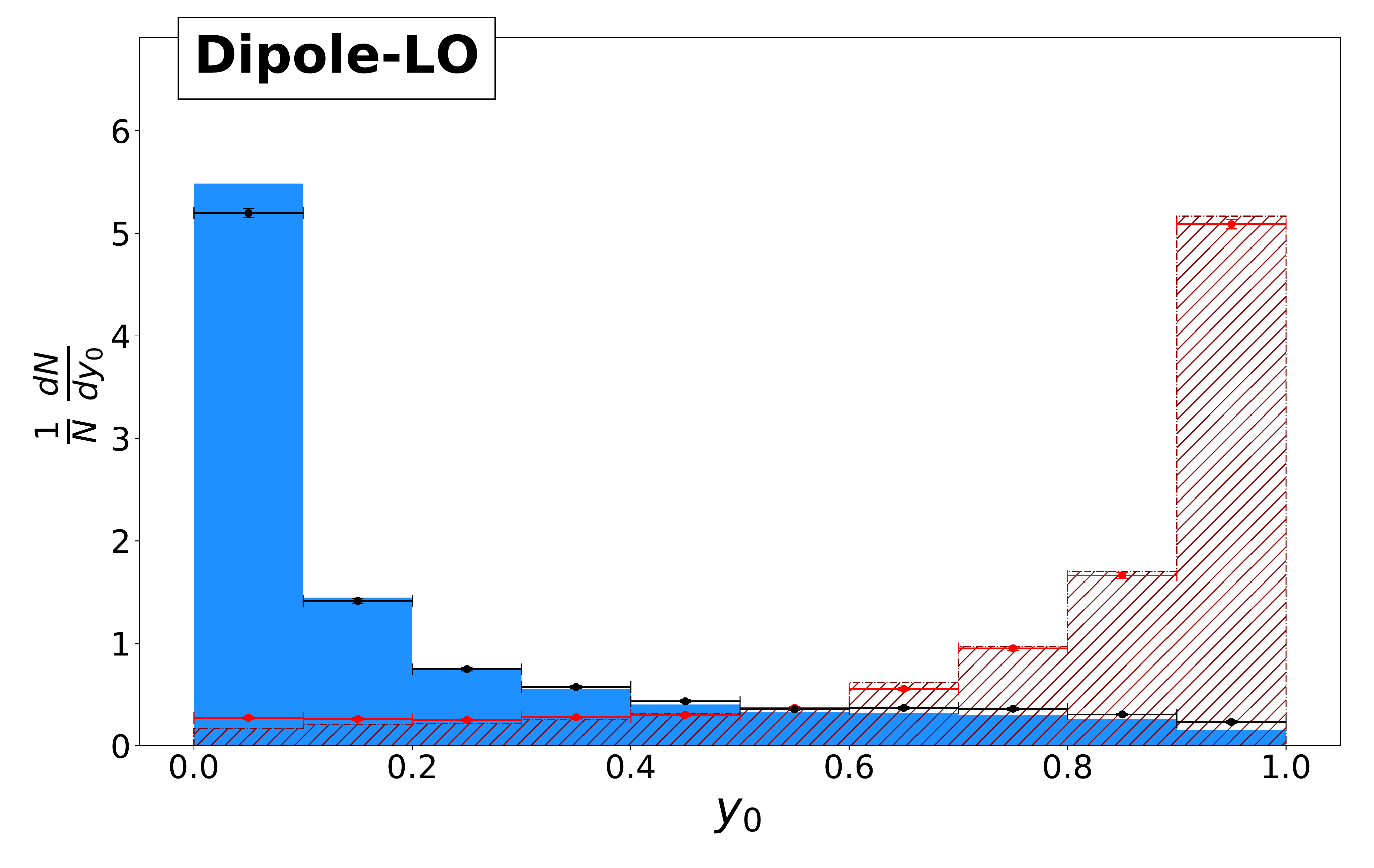}	
	\includegraphics[scale=0.18]{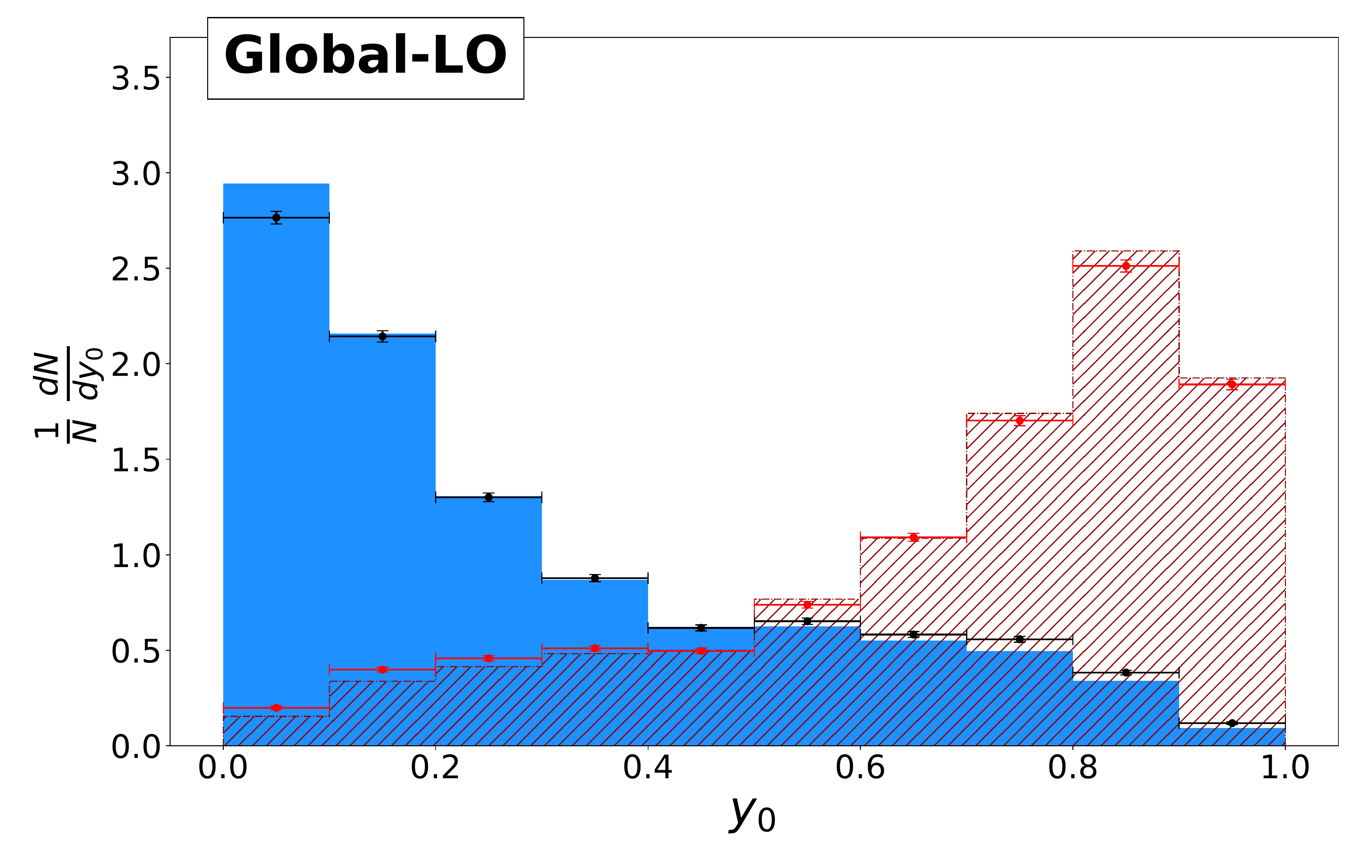}		
	\caption{Normalized binned distributions of the network output discriminating background from signal class for four different instances of signal simulations.}		
	\label{fig:nn_out}
\end{figure*} 					

In this section, we examine the performance of CNNs in identifying the different simulations of the same signal from the same background dataset described in Section~\ref{sec:data_sim}. When trained with the same architecture, the relative discrimination power should reflect the physical intuition we presented in the preceding section.  The four sets of signal events are preprocessed in the same manner as described in Section~\ref{sec:prepro} for the lower resolution, and the network $\mathcal{P}^{LR}_J$-CNN is trained with  the procedure described in Section~\ref{sec:net_train}. The performance on the higher resolution should follow the same trend, and hence unwarranted for the aim of the present work.  
\subsection{Effects of central radiation on the network output}

The two-dimensional histogram of the network output $y_0$ (the probability of an event being a signal) of the signal validation datasets with various variables quantifying the additional information beyond the two jet system are shown in Figure~\ref{fig:out_2d}. The weight of each event is set such that the total sum of all events with or without the third jet corresponds to one. Therefore, the total sum of the histogram with the physical quantities of the third jet corresponds to the fraction with at least one additional jet.\footnote{Due to the range of the variables, the total sum is not equal to the fraction presented in Section~\ref{sec:third_char}} The comparatively lower concentration of events for the Global-LO simulation is due to the lesser performance of the network (presented in Section~\ref{sec:res_perf}) compared to the other three simulations.  

In Figure~\ref{fig:out_pt}, where the histogram is with the transverse momentum of the third jet, we see that for the dipole recoil, both orders have the maximum concentration of events in the top left corner. The third jet has the least transverse momentum in this region, and the network identifies the event as most signal-like. For the case of the global recoil, we see that the NLO simulation has a higher concentration near the top-left corner. In contrast, the LO simulation has significantly reduced events near the top-left, with the shift towards the bottom in the y-axis more prominent. The greater change in the network output can be understood by recalling from Figure~\ref{fig:hist_2d} that the relative position of the third jet is much more central for the Global-LO simulation event if its transverse momentum is in a similar range. This property is further confirmed in Figure~\ref{fig:out_eta} where the histogram is on the $|\eta_{j_3}|$ and $y_0$ plane. The events for the Global-LO simulation is closer to the left side: implying that the third jets are much more central; and lower in the $y_0$ axis: indicating that the network identifies the signal less efficiently. Similarly, the same histogram for the dipole recoil scheme and different orders show a concentration of events in the top right corner, where the third jets are more forward, and the network identifies the signal with greater confidence.  

To look collectively into the events with or without a third jet, we define the scalar sum of $p_T$ between the two tagging jets as,
\begin{equation}
H^C_T=\sum_{\eta_i\in[\eta_l,\eta_u]}\; p_T^i\;,
\end{equation}
where the range is determined by the pseudorapidity of the two jets: $\eta_{j_1}$ and $\eta_{j_2}$ mapped such that $\eta_l<\eta_u$. We do not remove the particles within the jets when calculating $H_T^C$, thus giving a non-zero value for all events. As expected, we see in Figure~\ref{fig:out_htc}, that the Dipole-NLO simulation has the highest proportion of events near the top left corner, followed by Dipole-LO and Global-NLO, with Global-LO having a larger concentration in the central regions of the $(H^C_T,y_0)$-plane. Therefore, we see that events without the third jet also follow a similar pattern to those with the additional jet.

\begin{figure*}[t]
	\centering	
	\includegraphics[scale=0.19]{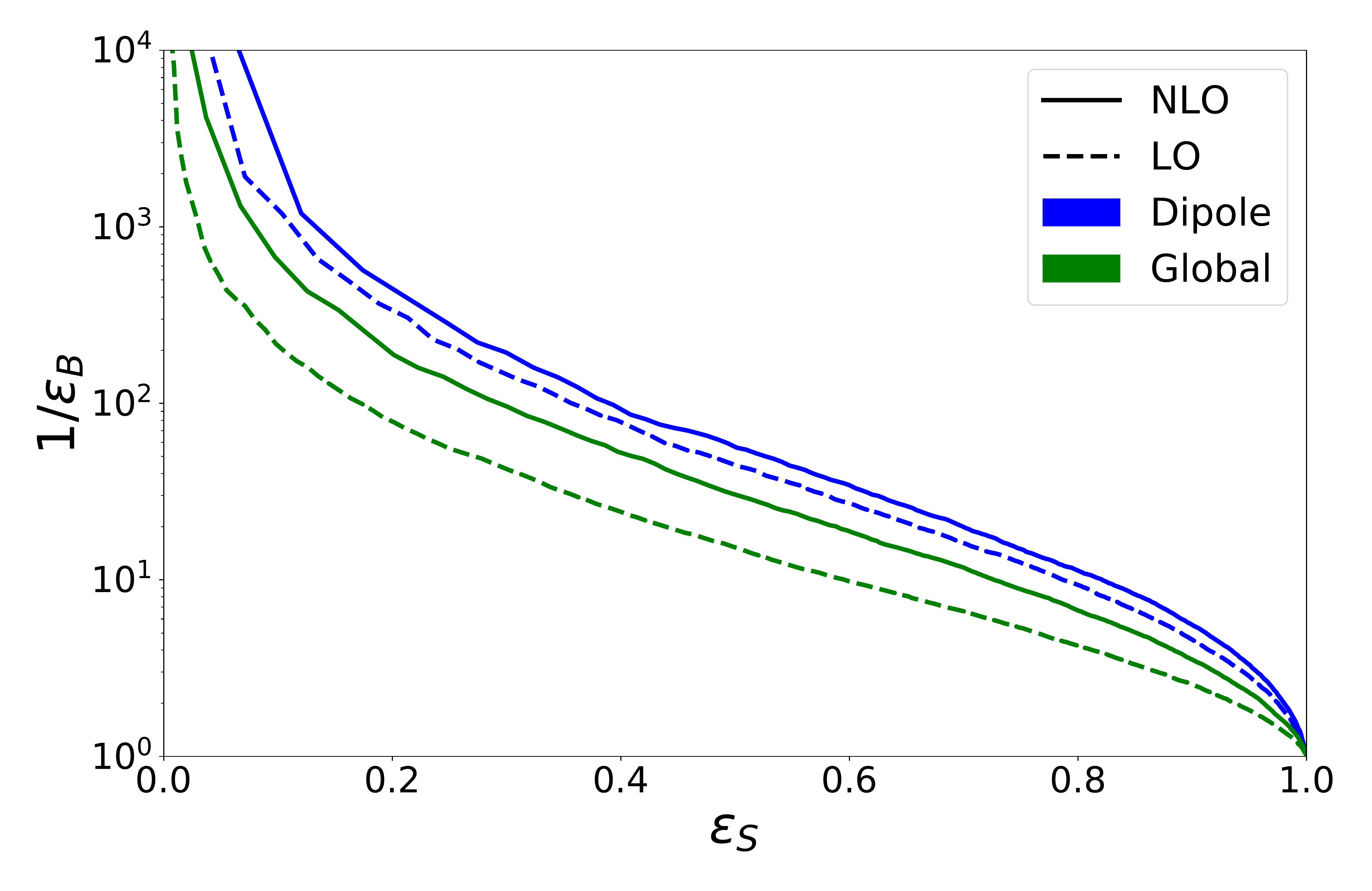} 
	\hspace{5mm}
	\includegraphics[scale=0.19]{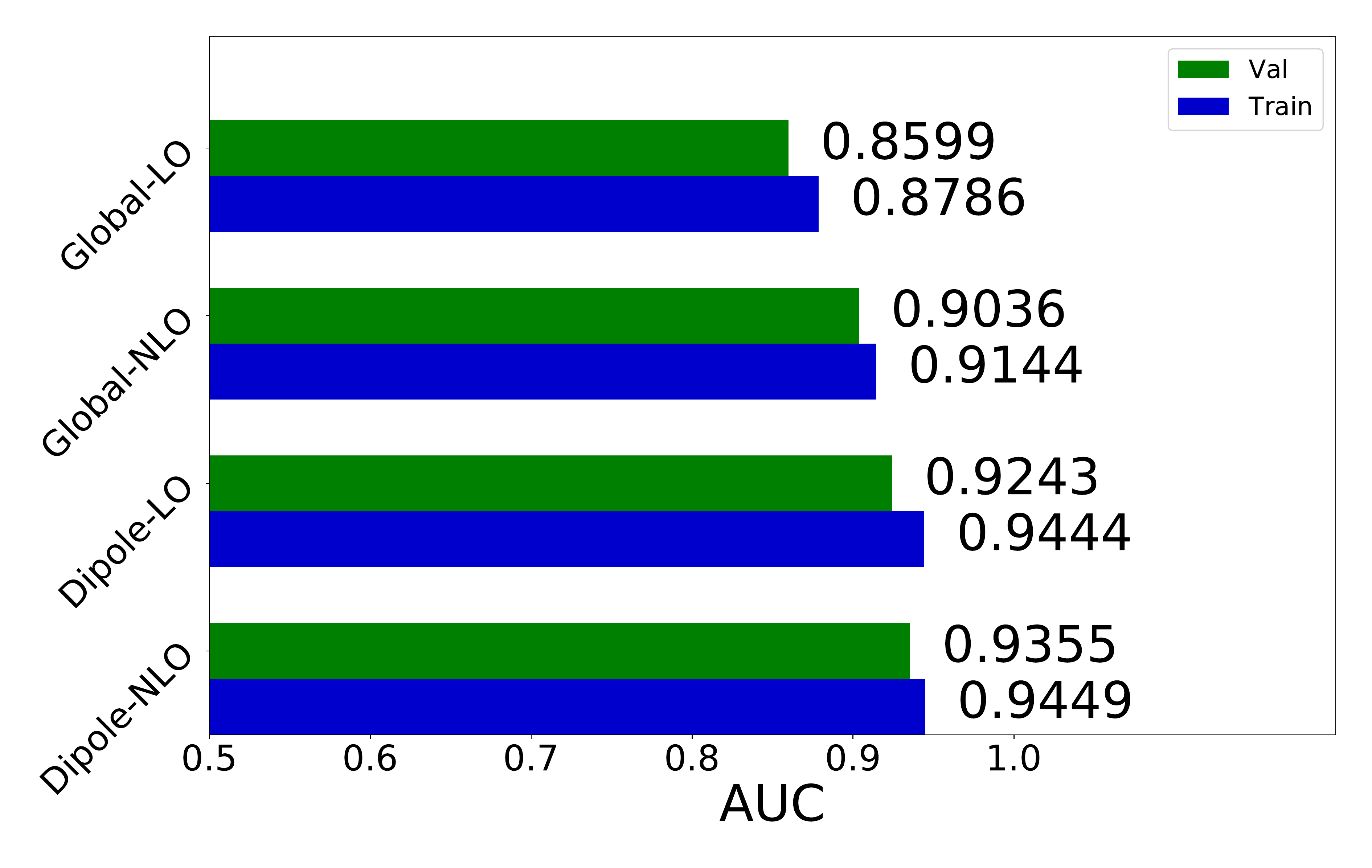}			
	\caption{Comparison of the performances in terms of Receiver operating characteristic (ROC) curves (left plot) on the validation dataset between the signal acceptance $\epsilon_S$ and the background rejection $1/\epsilon_B$, and the corresponding areas under these ROC curves (AUC) (right plot) for the training and validation data are shown for the four different cases of signal simulations.}		
	
	\label{fig:auc}
\end{figure*} 					

\subsection{Dependence of deep-learning performance on the signal simulation}
\label{sec:res_perf}

The normalized distribution of the network output $y_0$ of each class are shown in Figure \ref{fig:nn_out} for the four different signal simulation approaches. One can see that the CNN trained and validated with the Dipole-NLO simulation has the highest separation from the background.  To better quantify the power, we look at the Receiver operating characteristic (ROC) curves between the signal acceptance $\epsilon_S$ and the background rejection $1/\epsilon_B$, and the area under the ROC curve (AUC). These are shown in Figure \ref{fig:auc}. As expected, the highest discrimination is obtained for Dipole-NLO with a validation AUC of 0.9355, followed by Dipole-LO with 0.9243 validation AUC. Inadvertently, the Dipole-NLO signal happens to be the most physically accurate simulation. The hierarchy suggests that the recoil scheme is of greater importance than the perturbative accuracy for the CNN analysis with tower images. Looking at the global recoil for each order, we see that Global-NLO has better performance, with the CNN trained and validated with Global-LO having the least discriminatory power.  To understand this relative power, we note that the third jet in an NLO simulation has a leading-order accuracy. Whereas, for the LO case, the third jet, if present, is a consequence of the parton shower. The global-recoil scheme enhances the radiation in the central regions for both orders; however, it is partially controlled by the NLO simulation of the first real emmision, while there is no such control for the LO case.

Although we have seen that the network trained and tested with different signal simulations shows notable differences, it is worth investigating how a CNN trained on a specific simulation fare when tested on other signal simulations. The validation AUC for all signal simulations evaluated on each of the networks trained on the different signal simulations is shown in Table~\ref{tab:cross_auc}. For each signal type, the network it was trained on has the maximum discrimination, which is unsurprising given that the purpose of the training is to encode its behaviour into the network. Moreover, the trend of increasing performance is the same regardless of the signal dataset used in training, pointing towards all networks learning the underlying difference between the signal and the background. Another feature of interest is the relatively higher range of AUCs for the LO datasets than NLO ones, pointing towards their relatively high uncertainties. Interestingly, regardless of the nature of the simulation used during the training, the most accurate simulation amongst the four: Dipole-NLO events, have a very stable validation AUC with only a $1.6\%$ deviation. This stability shows that CNNs can learn the underlying differences between VBF events and non-VBF events even when the VBF simulation is suboptimal.

To gauge the possible improvement in using a dipole scheme over the global scheme used in our previous work, we train the CNN with the combined gluon-fusion signal and the instance of Dipole-NLO simulation of the VBF process in the same proportion as described in Section~\ref{sec:ML_VBF} and extract the bounds on the branching ratio. We find the median upper limit on the invisible branching ratio for an integrated luminosity $L=300$ fb$^{-1}$ to be 2.22\%.

	In all preceding analyses, we have used LO samples without any matching, and the third jet originates exclusively from the parton-shower, which is inaccurate in describing harder emissions. It is worth examining how a matching procedure between the hard matrix element and the parton shower, which improves the description of the third jet in the harder regions, influences the network performance. To inspect the possible improvement of such matching procedures, we generate VBF events matched with an additional jet via the MLM procedure~\cite{Mangano:2002,Mangano:2006rw} for both parton-recoil schemes. We found a continuous differential jet rate and transverse momentum distribution of the different jet samples for an {\tt xqcut} value of 100. As recommended for VBF processes, the {\tt auto\_ptj\_mjj} flag was set to false. All other aspects of the simulation including the renormalization and factorization scale, PDFs, and baseline selection criteria are the same as described in Section~\ref{sec:sig_gen}. We generated about 25k events after baseline selection for both recoil schemes. 
	Testing with these samples for the networks trained with the leading order unmatched samples with the same parton-shower recoil against the validation background dataset, we find an AUC of 0.8651 and 0.9261 for the global and dipole matched LO samples, respectively. Compared to the full NLO simulation values tested on these networks (Table~\ref{tab:cross_auc}), these values lie closer to the LO simulation, indicating that the matching procedure does not help alleviate the issues of the global parton shower. In contrast, the matched dipole value is still relatively stable, although closer to the LO value than the NLO value, signifying the relative importance of the virtual corrections of the NLO simulation.

\begin{table}[t!]
	\resizebox{0.6\textwidth}{!}{
		\renewcommand{\arraystretch}{1.4}
	\begin{tabular}{|c|l|c|c|c|c|}
		\hline 
		&\textbf{Train}&\multicolumn{4}{c|}{\textbf{Test Signal Dataset}}\\
		\cline{3-6}
		\textbf{Sl.No}&\textbf{Signal Dataset}&Globalo-LO&Global-NLO&Dipole-LO&Dipole-NLO\\
		\hline 
		1.&Global-LO&0.8599&0.8956&0.9027&0.9201\\
		2.&Global-NLO&0.8486&0.9036&0.9112&0.9288\\
		3.&Dipole-LO&0.8036&0.8878&0.9243&0.9335\\
		4.&Dipole-NLO&0.8234&0.8922&0.9200&0.9355\\\hline 
	\end{tabular}
}\caption{The table shows the test AUC evaluated for all signal simulation for each CNN trained on the different signal simulations. 
}\label{tab:cross_auc}
\end{table}


\section{Summary and conclusion}
\label{sec:conclusion}
The Large Hadron Collider, in its previous two runs, has already accumulated enough data to establish the Standard Model on a strong footing, pinpointing different properties of the Higgs boson and also setting strong constraints in diverse BSM scenarios. The vector boson fusion mechanism of production is unique in many of these Higgs measurements and BSM searches. The strongest upper limit on the Higgs's invisible branching ratio comes from this channel, although a significantly large window remains open above the SM prediction. An improved constraint can enormously squeeze the parameter space on many new physics scenarios, such as Higgs/scalar portal dark matter. 

In recent times advancements in different machine learning tools have opened the possibility to relook many of these analyses with sophisticated data-driven methods. Smartly designed neural networks demonstrate a capability to comprehend event kinematics and radiation pattern to a great extent, even when the event topology is rather simple. Invisible Higgs search in VBF is a process where the phenomenological study relies on two forward jets and a high missing transverse momentum. Moreover, many studies in this direction established some of the fundamental features of event shape, which has almost been a norm to control the extensive QCD and electroweak backgrounds. Our previous work took the invisible Higgs search in VBF as a case study. We constructed Convolutional Neural Networks (CNN) based deep-learning algorithms using just the low-level calorimeter inputs from the entire event topology without exclusive reconstructed objects. This novel approach can indeed provide the most stringent bounds on the invisible branching ratio of the SM-like Higgs boson, significantly outperforming the existing experimental search. 

It is evident that deep-learning algorithms with multiple non-linear hidden layers can efficiently characterize complex distribution functions describing the feature space with greater accuracy. This expressivity enhances their capability to distinguish the signal region from the background, exploiting maximal information, even if the event topology is relatively simple. However, to exploit the full potential of this extraordinary capability, a precise theoretical projection of the QCD parton shower and the corresponding radiation pattern is required. The present work demonstrates this interrelation utilizing different showering schemes with leading order and higher-order computation.

In this work, we carried out a quantitative analysis to investigate the dependence of a CNN's performance on the recoil scheme of the parton shower and the perturbative accuracy of the matrix element simulation for a VBF Higgs signal decaying to invisible particles. The difference between the leading order and next-to-leading order, although present, is not very
pronounced for the physically correct dipole-recoil scheme. We found that the training is highly dependent on the recoil scheme, with a better performance coming for the physically accurate dipole recoil. With this fortunate coincidence, a complete analysis with all VBF processes showered with a dipole recoil scheme will possibly reduce the upper limits on the invisible branching ratio even further than the projection which used a global recoil scheme.

\section*{Acknowledgement}
\label{sec:ack}
The work is supported by the Physical Research Laboratory (PRL), Department of Space, Government of India. Computations were performed using the HPC resources (Vikram-100 HPC) and TDP project at PRL. Authors gratefully thank Satyajit Seth and Aruna K Nayak for stimulating discussion and suggestions throughout this project. 
    
\appendix   
\section{Baseline Selection Criteria} 
\label{app:base_sel} 
We use the same selection criteria used in the deep-learning analysis of ref.~\cite{Ngairangbam:2020ksz}. This cuts except for the one on missing transverse energy were based on the experimental shape-analysis of ref.~\cite{CMS:2018yfx}. Here, we summarize them for completeness:  
\begin{itemize} 
	\item \textbf{Jet $p_T$}: at least two jets with the hardest(second-hardest) one having at least 80(40) GeV transverse momentum.
	\item \textbf{Lepton-veto:} Events should not have any reconstructed electron (muon) with minimum transverse momentum $p_T > 10$ GeV in withing the tracker region, {\sl i.e.} $|\eta|<2.5\,(2.4)$. 
	\item \textbf{Photon-veto:} Reject events having any photon with $p_T > 15$ GeV within $|\eta|< 2.5$.	
	\item \textbf{$\tau$ and b-veto}: If an event has any tau-tagged jets in $|\eta|<2.3$ with $p_T> 18$ GeV, or b-tagged jets in $|\eta|<2.5$ with $p_T>20$ GeV, they are discarded.
	\item \textbf{Minimum Missing transverse Energy $\slashed{E}_T$}: An event must have a minimum transverse energy, $\slashed{E}_T >200 $ GeV to be selected.
	\item \textbf{Alignment of MET with respect to jet directions}: The jets should have an azimuthal separation greater than 0.5 from $\slashed{E}_T$, {\sl i.e} $\min(\Delta \phi(\vec{p}_T^{\;\slashed{E}_T},\vec{p}_T^{\;j})) > 0.5$ for all jets upto the fourth leading jet satisfying $p_T > 30 $ GeV and $|\eta|< 4.7$, This requirement rejects QCD multijet backgrounds arising due to severe mismeasurement.
	\item \textbf{Jet rapidity}: The tagging jets should be well within the calorimeter acceptance($|\eta_j| < 4.7$), and at least one of them should be within the central regions ($|\eta_{j_i}| < 3$).
	\item \textbf{Jets in opposite hemisphere}: The tagging jets must reside in opposite hemisphere in $\eta$. This is acheived by imposing the condition $\eta_{j_1}\, \times \;\eta_{j_2}<0$. 
	\item \textbf{Azimuthal angle separation between jets}: We require the azimuthal seperation between the two tagging jets to satisfy $|\Delta \phi_{j_1j_2}|<1.5$.
	\item \textbf{Jet rapidity gap}: The rapidity gap between two leading jets must satisfy $|\Delta\eta_{j_1j_2}|>1$. 
	\item \textbf{Di-jet invariant mass}: The invariant mass of the two jet system should satisfy $m_{jj}>200$ GeV.
\end{itemize}  
After weighting the different background processes by their cross-sections and the baseline selection efficiency, we get almost $95\%$ contribution from non-VBF type production of $Z/W$ bosons while remaining is of the VBF origin. 

\bibliographystyle{JHEP}
\bibliography{ref}
\end{document}